\documentclass[
aip, 
apl, 
superscriptaddress,
amsmath, amssymb,
reprint,
floatfix
]{revtex4-1}

\usepackage[utf8]{inputenc}
\usepackage{hyperref}
\usepackage{graphicx}
\usepackage{textcomp}
\usepackage{color}
\usepackage{ulem}

\begin{document}

\title{Electric-field tuning of the valley splitting in silicon corner dots}
\date{\today}

\author{D. J. Ibberson}
\affiliation{Quantum Engineering Technology Labs, University of Bristol, Tyndall Avenue, Bristol, BS8 1FD, United Kingdom}
\affiliation{Hitachi Cambridge Laboratory, J. J. Thomson Ave., Cambridge, CB3 0HE, United Kingdom}
\author{L. Bourdet}
\affiliation{University Grenoble Alpes, CEA, INAC-MEM, 38000 Grenoble, France}
\author{J. C. Abadillo-Uriel}
\affiliation{Materials Science Factory, Instituto de Ciencia de Materiales de Madrid, ICMM-CSIC, Cantoblanco, E-28049 Madrid, Spain}
\author{I. Ahmed}
\affiliation{Cavendish Laboratory, University of Cambridge, J. J. Thomson Ave., Cambridge, CB3 0HE, United Kingdom}
\author{S. Barraud}
\affiliation{CEA/LETI-MINATEC, CEA-Grenoble, 38000 Grenoble, France}
\author{M. J. Calder\'{o}n}
\affiliation{Materials Science Factory, Instituto de Ciencia de Materiales de Madrid, ICMM-CSIC, Cantoblanco, E-28049 Madrid, Spain}
\author{Y-M. Niquet}
\affiliation{University Grenoble Alpes, CEA, INAC-MEM, 38000 Grenoble, France}
\author{M. F. Gonzalez-Zalba}\thanks{mg507@cam.ac.uk}
\affiliation{Hitachi Cambridge Laboratory, J. J. Thomson Ave., Cambridge, CB3 0HE, United Kingdom}

\begin{abstract}
We perform an excited state spectroscopy analysis of a silicon corner dot in a nanowire field-effect transistor to assess the electric field tunability of the valley splitting. First, we demonstrate a back-gate-controlled transition between a single quantum dot and a double quantum dot in parallel that allows tuning the device in to corner dot formation. We find a linear dependence of the valley splitting on back-gate voltage, from $880$~$\mu$eV to $610$~$\mu$eV with a slope of $-45\pm 3$~$\mu$eV/V (or equivalently a slope of $-48\pm 3$~$\mu$eV/(MV/m) with respect to the effective field). The experimental results are backed up by tight-binding simulations that include the effect of surface roughness, remote charges in the gate stack and discrete dopants in the channel. Our results demonstrate a way to electrically tune the valley splitting in silicon-on-insulator-based quantum dots, a requirement to achieve all-electrical manipulation of silicon spin qubits. 

\end{abstract}

\maketitle

A quantum bit implemented on the spin degree of freedom of a single electron in silicon is one of the most promising candidates for large-scale quantum computation due to its long coherence time~\cite{Tyryshkin2011}. Nowadays, single electrons spins can be confined in quantum dots (QDs)~\cite{Veldhorst2015} and single and two qubits operation can be performed with great accuracy~\cite{Zajac2018, Watson2018}. However, scaling to a large number of qubits remains a major challenge. Strategies at the architecture level propose large-scale quantum circuits~\cite{Vandersypen2017, Li2017}, using in particular CMOS technology for the implementation of error-correction protocols and the integration with classical electronics~\cite{Veldhorst2017,Schaal2018}. At the qubit level, all-electrical control of spins is desired because manipulation can be performed using local oscillating electric fields on gates that already define the QD. This is as opposed to magnetic-field-based qubit control that require microwave antennas or cavities that deliver less localized fields~\cite{Veldhorst2014}. All-electrical control of electron spins in silicon has been achieved using the extrinsic spin-orbit interaction (SOI) induced by magnetic field gradients from micromagnets~\cite{Kawakami2014,Yoneda2017}\, and recently a much more compact version has been demonstrated using the enhanced intrinsic SOI in silicon QDs with low-symmetry, such as CMOS corner dots~\cite{Corna17}. Silicon corner dots are distributed over two Si/SiO$_2$ interfaces and have a single symmetry plane~\cite{Voisin2014, Anasua2018}, as opposed to the more common planar silicon QDs~\cite{Angus2007a} that usually have two symmetry planes. The underlying control mechanism in the aforementioned study is based on the mixing of spin and valley degrees of freedom that allows driving electrically inter-valley spin rotations. A recent proposal suggests that the efficiency of this mechanism can be improved by tuning the valley splitting electrically to bring the qubit near the valley-mixing point for manipulation and away to mitigate decoherence. This opens an opportunity for compact electrical manipulation of spin qubits while retaining long coherence times~\cite{Bourdet2018}.

In this Letter, we demonstrate experimental control of the valley splitting in a silicon corner dot by means of static electric fields. We use a silicon-on-insulator nanowire field-effect transistor (NWFET) tuned to the corner dot regime and perform energy spectroscopy to quantify the valley splitting. We measure an electric field tunability of $-48\pm 3$~$\mu$eV/(MV/m) compatible with reports on planar silicon QDs~\cite{Yang2013, Veldhorst2014, Gamble2016}. The data are in agreement with tight-binding calculations that include the effect of surface roughness and remote Coulomb scattering. Our results are a basic ingredient for all-electrical spin qubit manipulation with tunable spin-valley mixing.

\begin{figure*}
	\centering
		\includegraphics{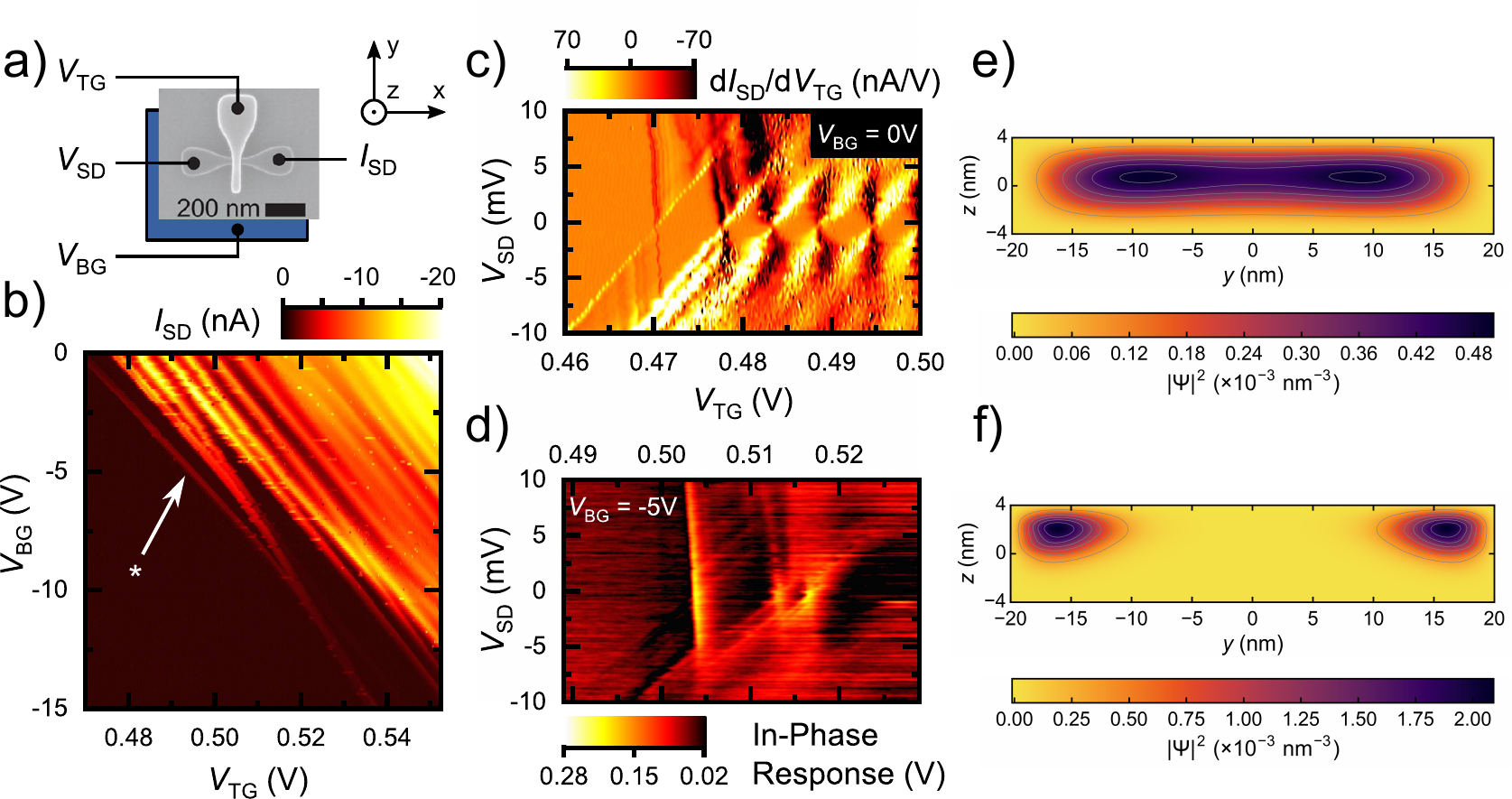}
	\caption{(a) Electron-microscope image of a device similar to the one measured, with the parameters indicating the measurement setup and the x,y,z axes. (b) Source-drain current as a function of top-gate ($V_\text{TG}$) and back-gate voltage ($V_\text{BG}$) at $V_\text{SD}=-2$~mV. The asterisk indicates the current oscillation used to perform excited state spectroscopy. (c) Coulomb diamonds taken at $V_\text{BG}=0$~V showing regularly-spaced oscillations (single dot regime). (d) Irregular Coulomb diamonds taken at $V_\text{BG}=-5$~V (double dot regime), measured in reflectometry. (e) Simulation of the electron probability distribution $\left|\Psi\right|^2$, in the y-z cross-section of the channel, when $V_\text{BG}=-1$~V, showing the formation of a single QD and (f) for $V_\text{BG}=-10$~V, illustrating the double QD regime.}
	\label{fig1}
\end{figure*}

We perform the experiments on a NWFET similar to the one pictured in Fig.~\ref{fig1}(a) at a refrigerator temperature of 40~mK. The device channel, oriented in the [110] direction, has a width of 42~nm, a height of 8~nm, and a gate length of 44~nm, and is doped with phosphorous at a concentration of $5 \times 10^{17}$ cm$^{-3}$. The gate oxide is formed of a SiO$_2$(0.8~nm)/HfSiON(1.9~nm) stack followed by a TiN(5~nm)/poly-Si(50~nm) metal gate. The undoped silicon substrate is activated by flashing a surface-mounted blue LED to generate free carriers, and it can then be used as a back gate.  Voltages can be applied to the top-gate ($V_\text{TG}$) and back-gate ($V_\text{BG})$ modifying the electrostatic potential in the channel, confining electrons in electrostatically-defined QDs.

We characterize the charge stability diagram of the device by measuring the source-drain current ($I_\text{SD}$) as a function of $V_\text{TG}$ and $V_\text{BG}$ in the sub-threshold regime of the transistor using a source-drain voltage $V_\text{SD}=-2~$mV, see Fig.~\ref{fig1}(b). We observe oscillations in $I_\text{SD}$ as a function of the gate voltages, characteristic of Coulomb blockade~ \cite{zwanenburg2013silicon, nagamune1994single}. At $V_\text{BG}=0$~V the current oscillations appear at regular $V_\text{TG}$ intervals. However, as $V_\text{BG}$ is reduced, the oscillations become more irregular and we even observe regions where these oscillations anticross (see $(V_\text{TG},V_\text{BG})=(0.49,-1)~$V and $(0.53,-8)~$V for example). The stability diagram points towards a transition from a single QD regime at high $V_\text{BG}$ to a double QD in parallel at low $V_\text{BG}$ which we investigate further in Fig.~\ref{fig1}(c,d). In panel (c), we plot the differential conductance as a function of $V_\text{SD}$ and $V_\text{TG}$ at $V_\text{BG}=0~$V. We see diamond-shaped regions of zero differential conductance, i.e. Coulomb diamonds. The first diamond, of larger size, indicates that as the dot is depleted, its size is reduced and hence its addition energy increases. This pattern is characteristic of transport through a single few-electron quantum dot~\cite{Kouwenhoven2001,Fuechsle2010}. In this regime, we measure 6~meV for the first addition energy and 3.75~meV for consecutive ones. We simulate the electron probability distribution in the channel of the device in Fig.~\ref{fig1}(e) (see details later). Our calculations confirm the formation of a single QD whose electron probability distribution $\left|\Psi\right|^2$ extends over the central part of the channel.

Next, in Fig.~\ref{fig1}(d), we measure the Coulomb diamonds at $V_\text{BG}=-5$~V, now using gate-based radio-frequency reflectometry~\cite{Colless2013, Gonzalez-Zalba2015}. In this case, we plot the in-phase response of the resonator, which shows an enhancement at the regions of charge instability, i.e. at the edges of the Coulomb diamonds. The plot reveals diamonds whose size changes non-monotonically with increasing electron number. These findings combined with the anticrossings observed in Fig.~\ref{fig1}(b) indicate that a double QD in parallel has formed in the channel. To back up our conclusion, we also simulate the electron probability distribution under these new bias conditions in Fig.~\ref{fig1}(f). Now, the electron probability is distributed over two regions located at the top-most corners, i.e. silicon corner dots~\cite{Voisin2014}. 

\begin{figure}
	\centering
		\includegraphics{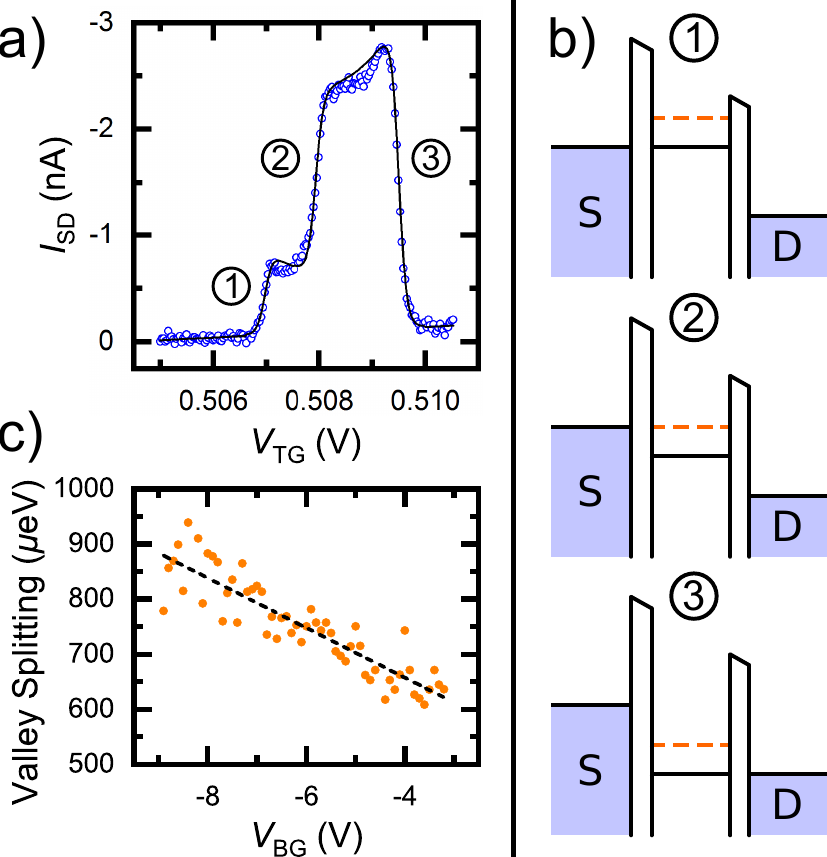}
	\caption{(a) First Coulomb oscillation at $V_\text{BG}=-7.8$~V, taken from the data in Fig.~\ref{fig1}(b) (transition marked with an asterisk). Data (hollow blue dots) and simulated source-drain current (solid black line). (b) Diagrams illustrating the respective alignment of the QD and source and drain electrochemical levels at the three features observed in (a). Top panel: GS, in solid black, and source aligned. Middle panel: ES, in dashed orange, and source aligned. Bottom panel: GS and drain aligned. (c) Plot of valley splitting against $V_\text{BG}$ (orange dots) with a best-fit line gradient of $-45\pm 3$~$\mu$eV/V (black dashed line).}
	\label{fig2}
\end{figure}

Having recognized the formation of double corner dots, we move on to the study of the energy spectrum of the first electronic transition of one corner dot observed in Fig.~\ref{fig1}(b), marked with an asterisk. We restrict our analysis to back-gate voltages in the $[-9, -3]$~V range where the corner QDs form. At $V_\text{BG}<-9~$V the first electron transition of both dots start to overlap making the excited state analysis ambiguous. At even lower back-gate voltages $V_\text{BG}<-12$~V only one of these two Coulomb oscillations is visible indicating that one of the tunnel barriers of one of the QDs has become too opaque to observe transport through. In Fig.~\ref{fig2}(a), we plot $I_\text{SD}$ as a function of $V_\text{TG}$ at $V_\text{SD}=-2~$mV and $V_\text{BG}=-7.8~$V for the first oscillation in Fig.~\ref{fig1}(b). Here, it is possible to resolve three distinct components, which are explained in Fig.~\ref{fig2}(b). The left shoulder corresponds to alignment of the source and dot ground state (GS) electrochemical potentials, where conduction from source to drain via the QD's GS becomes possible. At the middle feature, an excited state (ES) is aligned with the source, providing an additional conductance channel through the dot~\cite{escott2010resonant}. We identify this ES as a valley excitation of the dot, as we shall see later. Finally, the right shoulder corresponds to alignment of the dot and drain electrochemical potentials. By increasing $V_\text{TG}$ further, we re-enter the Coulomb blockade regime. The data can be reproduced by considering the system in the sequential multi-level transport regime~\cite{zwanenburg2013silicon}. We assume fast relaxation from the ES in to the GS such that electrons always exit the QD via the GS. In this regime, $I_\text{SD}=e(\gamma_\text{in}^\text{GS}+\gamma_\text{in}^\text{ES})\gamma_\text{out}^\text{GS}/(\gamma_\text{in}^\text{GS}+\gamma_\text{in}^\text{ES}+\gamma_\text{out}^\text{GS})$ where $\gamma_\text{in}^\text{GS(ES)}$ correspond to the tunnel rate from the source reservoir to the GS(ES) of the QD and $\gamma_\text{out}^\text{GS}$ is the tunnel rate from the GS to the drain reservoir. We consider discrete states in the QD (0D density of states) and 1D density of states in the reservoirs. We plot the simulated current as a black solid line in Fig.~\ref{fig2}(a) which allows us to obtain the $V_\text{TG}$ values at which the electrochemical levels align. The position of the three components were tracked in $V_\text{TG}$ space over the range $-3 \geq V_\text{BG} \geq -9$ V, and the energy of the valley splitting calculated from these is plotted in Fig.~\ref{fig2}(c). The valley splitting varies from $880~\mu$eV at $V_\text{BG}=-9~$V to $610~\mu$eV at $V_\text{BG}=-3~$V, and decreases linearly in-between, with a slope of $-45 \pm 3~\mu$eV/V. These values are within the range observed for planar metal-oxide-semiconductor QDs~\cite{lim2011spin,kobayashi2016resonant} and similar to previously reported for silicon corner dots~\cite{Voisin2014}. The electron temperature in our measurements is 140 mK, limiting the minimum resolvable valley splitting to $40~\mu$eV. However, as we shall see later, this value is below the valley splitting expected for realistic sample configurations and well below the value reported above. The magnitude of the measured splitting is incompatible with the large valley splitting, of the order of several meV, observed for single dopants~\cite{Roche2012}.

\begin{figure}
	\centering
		\includegraphics[width=0.5\textwidth]{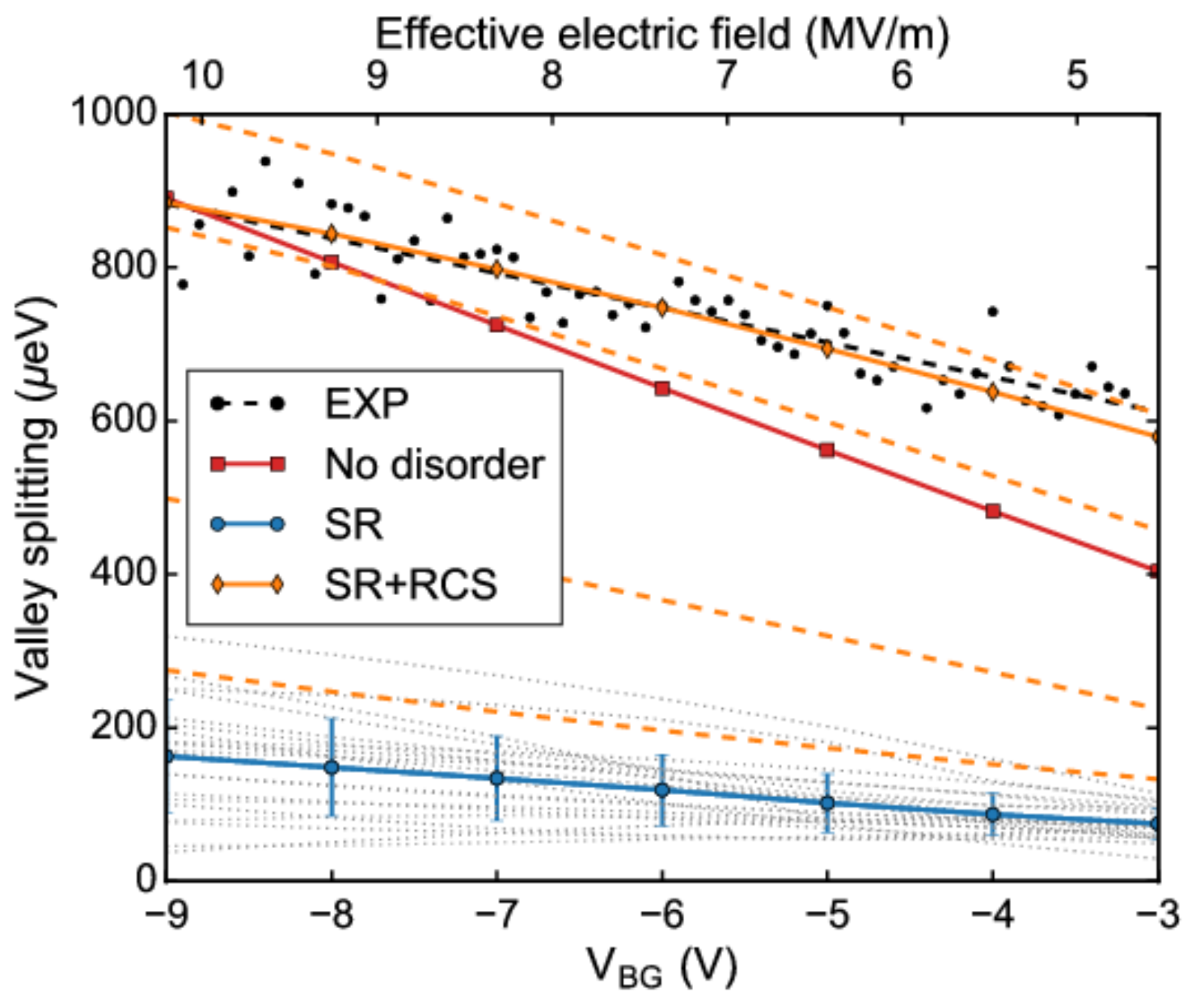}
	\caption{Simulations of the valley splitting as a function of $V_\text{BG}$. The red line gives the trend for a perfect device with no defects or roughness. The blue line and error bars are the average and standard deviation for different SR profiles with rms 0.35~nm (each plotted as a dashed gray line). Finally, the orange dashed lines are a few representative simulations with SR and RCS included (the best match with the experiment being highlighted by the solid orange line with diamond symbols). The top axis indicates the effective electric field perpendicular to the substrate (calculated in the reference, perfect device). The smaller in-plane effective field has little effect the valley splitting.}
	\label{fig3}
\end{figure}

In order to back up this interpretation, we have calculated the electronic structure of the device with a $sp^3d^5s^*$ tight-binding (TB) model~\cite{Niquet09}. The TB method describes all valleys at once and, therefore, captures their interactions in a rapidly varying potential. The single-particle energy levels are computed in the electrostatic potential from the gates (see Refs. \cite{Corna17,Bourdet2018} for details). The surface of the channel is passivated with pseudo-hydrogen atoms, but the effective oxide model of Ref.~\onlinecite{Kim11} gives similar trends. The simulations reproduce the single to double dot transition, as shown in Fig.~\ref{fig1}. The TB valley splitting of a single corner dot is plotted as a function of $V_\text{BG}$ in an ``ideal'' device without disorder in Fig.~\ref{fig3} (red line). The simulation reproduces the general features of the data since the valley splitting increases with negative back gate voltages as the wave function gets further pushed at the Si/SiO$_2$ interface. However, the calculated valley splitting is lower than the experiment at small $V_\text{BG}$ and shows a larger slope. Yet higher-lying orbital excitations in the same valley are significantly farther above, and can hardly explain the experimental results. The difference arises from ``imperfections'' of the device.

The magnitude of the valley splitting depends strongly on disorder -- especially surface disorder. On the one hand, surface roughness (SR) generally decreases the valley splitting~\cite{Culcer10} due to the interferences between reflections on different atomic planes. On the other hand, donors in the channel or traps in the oxide can increase it. However, donors can be ruled out as the cause of the enhancement since the slope of the valley splitting of donor-bound states is expected to be opposite to the experimental one, as a negative back gate voltage tends to push the electron away from the dopant~\cite{Rahman2011, Roche2012}. Coulomb disorder in the gate stack may increase the valley splitting by enhancing the electric field locally. Room-temperature mobility measurements in high-$\kappa$/metal gate devices actually show the fingerprints of ``Remote Coulomb Scattering'' (RCS~\cite{Casse06}) by charges at the SiO$_2$/HfSiON interface~\cite{Bourdet2016,Zeng2017b} with apparent densities as large as a few $10^{13}$ cm$^{-2}$. In fact, the Coulomb disorder in the gate stack likely results from a combination of charge traps at this interface, local band offset fluctuations (interface dipoles), and possibly from work function fluctuations in granular metal gates~\cite{Zeng2017a}. The existence of significant disorder is consistent with the fact that the two corner dots are not fully symmetric, as revealed by the stability map in Fig. \ref{fig1}(b) and the Coulomb diamonds in Fig. \ref{fig1}(d).

 %On the other hand, donors in the channel increase it (possibly up to about 10-20 meV~\cite{Roche2012}) due to the steep potential around the impurity. However, the slope of the valley splitting of donor-bound states is expected to be opposite to the experimental one, as a negative back gate voltage tends to push the electron away from the dopant~\cite{Rahman2011}. Coulomb disorder in the gate stack may also increase the valley splitting by enhancing the electric field locally. Room-temperature mobility measurements in high-$\kappa$/metal gate devices actually show the fingerprints of ``Remote Coulomb Scattering'' (RCS~\cite{Casse06}) by charges at the SiO$_2$/HfSiON interface~\cite{Bourdet2016,Zeng2017b} with apparent densities as large as a few $10^{13}$ cm$^{-2}$. In fact, the Coulomb disorder in the gate stack likely results from a combination of charge traps at this interface, local band offset fluctuations (interface dipoles), and possibly from work function fluctuations in granular metal gates~\cite{Zeng2017a}. The existence of significant disorder is consistent with the fact that the two corner dots are not fully symmetric, as revealed by the stability map in Fig. \ref{fig1}(b) and the Coulomb diamonds in Fig. \ref{fig1}(d).}

We have introduced roughness in the TB calculations using a Gaussian auto-correlation function~\cite{Goodnick85} for the Si/SiO$_2$ interface profile with correlation length $L_c=1.5$ nm and rms $\Delta_{\rm SR}=0.35$ nm adjusted~\cite{Zeng2017b} on room-temperature mobility measurements in similar devices. $\Delta_{\rm SR}$ lies in the upper range of the values usually reported for Si/SiO$_2$ interfaces, which is not unexpected for an etched device. As discussed above, the SR consideraly decreases the valley splitting (see Fig. \ref{fig3}) and is responsible for a significant device-to-device variability -- however the present experimental data are more than five standard deviations away from the calculations. The introduction of Coulomb disorder in the gate stack can raise the valley splitting back into the experimental range. Here we have modeled this disorder as a distribution of positive and negative charges at the SiO$_2$/HfSiON with total density $\sigma=10^{13}$ cm$^{-2}$. This large value is, again, consistent with the mobilities measured in planar as well as nanowire devices, and with their dependence on the thickness of the SiO$_2$ layer~\cite{Zeng2017b,Zeng2017a}. It must be seen as an effective density of charges mimicking all high-$\kappa$/metal gate-related disorders described above, which have similar fingerprints on the potential in silicon.

We point out, though, that the magnitude of the valley splitting depends strongly on the position of these charges. The localization is, indeed, much more efficient in a Coulomb than in a short-range SR potential, but also much more variable. For a given density of RCS charges, the valley splitting spans about one order of magnitude depending on their distribution. A statistical analysis of both mechanisms shows that 20 out of 20 simulated rough devices show well defined corner states at negative $V_\text{BG}$, while only 14 out of 20 simulated devices with RCS included still do so. Coulomb disorder must, therefore, primarily be reduced in order to mitigate device variability. As a matter of fact, the valley splitting has been measured in a similar device with two corner dots in parallel but with only SiO$_2$ as the gate dielectrics~\cite{betz2015dispersively}. The valley splitting at $V_\text{BG}=-1$ V was found to be 145 $\mu$eV in one dot, which is more compatible with the TB valley splitting calculated with SR and no RCS. This calls for a careful assessment of the sources of disorder in silicon devices. Removing high-$\kappa$ oxides from the gate stack might help to reduce Coulomb disorder and variability, at the price of a lower gate coupling.

In conclusion, we have highlighted a transition from single to double quantum dot (corner dots) in a silicon NWFET. We have also demonstrated that the valley splitting in one of the corner dots could be tuned from $880$~$\mu$eV to $610$~$\mu$eV by varying the static gate voltages (with a gradient of $-48\pm 3$~$\mu$eV/(MV/m) with respect to the effective field). The magnitude of the valley splitting and its dependence on the electric field can be reproduced by a tight-binding model when accounting for surface roughness and charges trapped in the gate oxides. Our results fulfill a milestone towards all-electrical spin manipulation using tunable valley-spin mixing. 

This research has received funding from the European Union's Horizon 2020 Research and Innovation Programme under grant agreement No 688539 (http://mos-quito.eu) and the Winton Programme of the Physics of Sustainability. DJI is supported by the Bristol Quantum Engineering Centre for
Doctoral Training, EPSRC grant EP/L015730/1. JCAU and MJC acknowledge funding from MINEICO (Spain) and FEDER via Grant FIS2015-64654-P. JCAU thanks the support from grant BES-2013-065888. IA is supported by the Cambridge Trust and the Islamic Development Bank.

\bibliography{references}
\end{document}